\RequirePackage{fix-cm}
\documentclass[twocolumn,epjc3]{svjour3}
\smartqed  
 \RequirePackage{mathptmx}      
\RequirePackage{graphicx}
\RequirePackage{lipsum}
\RequirePackage{mathtools}
\RequirePackage{cuted}
\RequirePackage{mathptmx}      
\RequirePackage{flushend}
\RequirePackage{latexsym}
\RequirePackage[numbers,sort&compress]{natbib}
\RequirePackage[dvipdfmx]{hyperref}
\journalname{
}
\begin{document}

\title{Noncommutativity of four-dimensional axisymmetric spacetime in polar coordinate
}


\author{Ryouta Matsuyama\thanksref{e1,addr1}
        \and
       Michiyasu Nagasawa\thanksref{addr1,addr2} 
}

\thankstext{e1}{e-mail: r201770192ve@jindai.jp}


\institute{Department of Physics, Graduate School of Science, Kanagawa University, Kanagawa 259-1293, Japan  \label{addr1}
           \and
           Department of Mathematics and Physics, Faculty of Science ,Kanagawa University, Kanagawa 259-1293, Japan \label{addr2}
}

\date{
}

\maketitle

\begin{abstract}
It is shown that in the noncommutative spacetime defined by the generalized Moyal product, consistent noncommutativity can be obtained independent of the coordinate system such as Cartesian or polar one.
In addition, based on the fact that the generalized Moyal product can be applied to arbitrary spacetime with non-trivial curvature, the effect of noncommutativity in the axisymmetric spacetime with central mass is investigated.
The results demonstrate how the noncommutativity of spacetime modify the shape of the stationary limit surface in the noncommutative Kerr spacetime, which implies that the gravitational interaction seems to be effectively softened.
\end{abstract}

\section{Introduction}
\label{intro}
The noncommutative spacetime first proposed by Snyder in the 1940s \cite{1} has once again become an important research topic in modern physics.
The reasons are that noncommutative spacetime naturally appears  from string theory \cite{2} and noncommutative geometry would be the basis of quantum theory \cite{3,4}.
Recently, several quantum gravitational theories based on spacetime noncommutativity have been proposed \cite{16,19,20,21,22,25,30,31,32,33,34}, and in the future they are expected to be applied to physical phenomena in regions where quantum effects are prominent, such as in the early universe and around black holes.
However, at present, such applications of noncommutative gravity have not progressed so much, due in part to the difficulty of establishing consistent noncommutative spacetime description by an arbitrary coordinate selection.
In the first place, even the polar coordinate system usually employed in gravity research has only recently begun to be studied in the field of noncommutativity \cite{5,6,7,8}.

Noncommutative spacetime is characterized by the nontrivial commutation relation $[{\hat x}^\mu,{\hat x}^\nu ]=i\theta^{\mu\nu}$, which is usually defined in the Cartesian coordinate system. 
For example, it is represented as $[{\hat x}, {\hat y}]=i\Theta$ in a two-dimensional space where $({\hat x}, {\hat y})$ are coordinate operators and $\Theta$ is a constant noncommutative parameter.
On the other hand, there is no clear form of commutation relation on the polar coordinate system, although some proposals have been made in previous works.
In \cite{5}, in order to investigate BTZ black hole in three-dimensional anti-de Sitter spacetime, the commutation relation $[{\hat r}, {\hat \phi}] = i\Theta {\hat r}^{-1}$ was employed.
In \cite{6}, the similar relation is derived from defining ${\hat x} = {\hat r} \cos {\hat \theta}$, ${\hat y} = {\hat r} \sin {\hat \theta}$.
Furthermore, in \cite {7,8} using the Moyal deformation quantization \cite{10}, higher order term of commutation relation between real space coordinates $r = \sqrt{x^2 + y^2}$ and $\phi=\arctan(y/x)$ by the expansion of $\Theta$ was calculated.
In particular, in \cite{8}, the behavior of the commutation relation in the limit when $r \rightarrow \infty$ and $r \ll \Theta$ was investigated by deformation quantization and Borel resummation.

In any coordinate system, the original noncommutativity is hard to be seen in most cases as long as the methods described in the above paragraph are used.
Most importantly, previous works have derived the commutation relation in the polar coordinate by way of the Moyal product defined in the Cartesian coordinate.
In other words, these methods have focused only on the polar coordinate in two-dimensional flat space and are not suitable for applied research in more than three-dimensional and curved spacetime.
In particular, in gravitational theory, the polar coordinate in four-dimensional spacetime are generally used in cosmology and black hole analysis, so that the detailed understanding of the strong gravity effect will not proceed unless we find out other framework than the transformation from the Cartesian coordinate.

The generalized Moyal product has been redefined as a function of coordinate using vielbein in order to realize the dynamical noncommutativity in scalar field theory in flat noncommutative spacetime \cite{12,13,14}.
In this paper, we showed that the generalized Moyal product gives consistent commutation relations in Cartesian and polar coordinate.
In the foregoing studies of polar noncommutativity \cite{5,6,7,8}, only a commutation relation of two-dimensional space was derived, but we have specifically dealt with a commutation relation in four-dimensional spacetime.
Furthermore, since the generalized Moyal product is available even in arbitrary spacetime, we investigated the effect of noncommutativity in the noncommutative Kerr spacetime.
As a result, we have found that  the stationary limit surface becomes more oblate as if the effect of gravity is alleviated in the noncommutative spacetime, which is consistent with the achievements of the literature \cite{16, 23, 24}.

The paper is organized as follows.
Section \ref{sec2} describes the generalized Moyal product.
Section \ref{sec3} shows nontrivial commutation relation in the polar coordinate system can be established using the generalized Moyal product.
Section \ref{sec4} examines the noncommutativity of spacetime in the Kerr metric.
Finally, we conclude this paper in Section \ref{sec7}.
Throughout this paper, we use the geometrical unit, $G=c=1$.


\section{\label{sec2}Generalized Moyal product}
One of the promising noncommutative spacetime construction methods is the deformation quantization by the generalized Moyal product $\star$, under which the product of arbitrary functions $f(x)$ and $g(x)$ is expressed as 
\begin{eqnarray}
\label{101}
f \star g 
= \exp \bigg[ \frac{i}{2}\theta^{\mu\nu}(x) \frac{\partial}{\partial x^\mu} \frac{\partial}{\partial y^\nu}\bigg] f(x)g(y) \bigg|_{x=y} ,
\end{eqnarray}
where $\theta^{\mu\nu}(x)$ is an antisymmetric tensor.
Furthermore, when expanding (\ref{101}) with respect to $\theta^{\mu\nu}$ until the second order, it coincides with the form of the Kontsevich product \cite{15,17}
\begin{eqnarray}
\nonumber
f \star g = fg &+& \frac{i}{2}\theta^{\mu\nu}\partial_\mu f \partial_\nu g - \frac{1}{8} \theta^{\mu\nu}\theta^{\rho\sigma}\partial_\mu \partial_\rho f \partial_\nu \partial_\sigma g \\
\label{1011}
&-& \frac{1}{12}\theta^{\mu\nu}\partial_\nu \theta^{\rho\sigma}(\partial_\mu \partial_\rho f \partial_\sigma g - \partial_\rho f \partial_\mu \partial_\sigma g),
\end{eqnarray}
which we use hereafter.
Under (\ref{101}), the nontrivial commutation relation of coordinates holds as
\begin{eqnarray}
\label{102}
[x^\mu, x^\nu]_\star = x^\mu \star x^\nu - x^\nu \star x^\mu = i\theta^{\mu\nu}.
\end{eqnarray}
In this paper, in order to focus on noncommutativity in arbitrary coordinate systems, $\theta^{\mu\nu}$ is defined as
\begin{eqnarray}
\label{103}
\theta^{\mu\nu}(x) = \theta^{ab}e_a{^{\mu}}(x)e_b{^{\nu}}(x),
\end{eqnarray}
according to the previous research\cite{12,13,14}.
$e_a{^{\mu}}$ and $e^a{_{\mu}}$ are vielbein, and $\theta^{ab}$ is a constant antisymmetric tensor defined on the tangent space.
As usual, the Greek alphabet ($\mu, \nu, \dots$) denotes indices related to spacetime and the lower case alphabet ($a, b, \dots$) are used to represent tangent space.
Note that $\theta^{ab}$ is the constant tensor in tangent space in contrast to $\theta^{\mu\nu}(x)$ which is the function of $x^\mu$ in spacetime.
Vielbein connects the tangent space defined by Minkowski metric $\eta_{ab}=(-1, +1, \cdots, +1)$ with the spacetime defined by classical metric tensor $g_{\mu\nu}$.
This role of vielbein is clearly recognized from its definition as 
\begin{eqnarray}
\label{104}
g_{\mu\nu} = \eta_{ab} e^a{_{\mu}}e^b{_{\nu}}, \hspace{5mm} \eta_{ab} = g_{\mu\nu} e_a{^{\mu}}e_b{^{\nu}}.
\end{eqnarray}
Also, vielbein satisfies orthonormality as
\begin{eqnarray}
\label{105}
e^a{_{\mu}} e_a{^{\nu}} = \delta^\nu_\mu, \hspace{5mm} e^a{_{\mu}} e_b{^{\mu}} = \delta^a_b.
\end{eqnarray}
In any coordinate system transformation $x^\mu \rightarrow x'^{\alpha} $, by choosing the corresponding vielbein appropriately, the Moyal product is defined as 
\begin{eqnarray}
\label{108}
\exp \bigg[ \frac{i}{2}\theta^{\mu\nu}(x) \frac{\partial}{\partial x^\mu} \frac{\partial}{\partial x^\nu}\bigg] = \exp \bigg[ \frac{i}{2}\theta'^{\alpha\beta}(x') \frac{\partial}{\partial x'^\alpha} \frac{\partial}{\partial x'^\beta}\bigg],
\end{eqnarray}
without any contradiction.
Due to this property, consistent noncommutativity can be defined independent of coordinate systems.
Moreover, it is assumed that each component of $\theta^{ab}$ in tangent space should be
\begin{eqnarray}
\label{106}
\theta^{ab} = \left(
\begin{array}{cccc}
0 & \Sigma & \Sigma & \Sigma \\
-\Sigma & 0 & \Theta & -\Theta \\
-\Sigma & -\Theta & 0 & \Theta \\
-\Sigma & \Theta & -\Theta & 0 \\
\end{array} 
\right),
\
\end{eqnarray}
when we deal with four-dimensional spacetime, so that $\theta^{ab}$ represents homogeneous noncommutativity of space.
Since vielbein $e_a{^{\mu}}$ is equal to  ${\rm diag} (+1,+1, +1, +1)$ in Cartesian coordinate in a flat spacetime, the commutation relation of coordinates is equivalent to the conventional one such as
\begin{eqnarray}
\label{107}
\begin{split}
[t,x]_\star = [t,y]_\star = [t,z]_\star = i\Sigma, \\
[x,y]_\star = [y,z]_\star = [z,x]_\star = i\Theta,
\end{split}
\end{eqnarray}
in this case.

\section{\label{sec3}Noncommutative polar coordinate}
The nontrivial commutation relation, which is essential in constructing noncommutative spacetime, must of course be well-defined when coordinate transformations is applied.
In this section, we confirm that noncommutativity of spacetime in the polar coordinate system defined by the generalized Moyal product is identical to that in the Cartesian coordinate system when compared at a suitable position.

The vielbein in polar coordinate $(t, r, \phi)$ of a three dimensional spacetime are
\begin{eqnarray}
\label{4.1}
e^a{_{\mu}} = \left( \hspace{-0.5mm}
\begin{array}{ccc}
  1 & 0 & 0 \\
  0 & \cos\phi & -r\sin\phi  \\
  0 & \sin\phi & r\cos\phi  \\
\end{array} 
\hspace{-1.5mm}\right) \hspace{-1mm},
e_a{^{\mu}} = \left(\hspace{-0.5mm}
\begin{array}{ccc}
  1 & 0 & 0 \\
  0 & \cos\phi & -\sin\phi /r  \\
  0 & \sin\phi & \cos\phi /r  \\
\end{array} 
\hspace{-1.5mm}\right).
\
\end{eqnarray}
Throughout this paper, the good triads and tetrads which appeared in a previous study \cite{40} are used for vielbein.
From the Moyal product corresponding to the polar coordinate defined by (\ref{4.1}) and (\ref{1011})--(\ref{103}), the commutation relation between any functions $f(t, r,\phi)$ and $g(t, r,\phi)$ can be derived as
\begin{eqnarray}
\nonumber
[f,g]_\star &=& i\Sigma(\cos\phi + \sin\phi)\bigg( \frac{\partial f}{\partial t} \frac{\partial g}{\partial r} - \frac{\partial f}{\partial r} \frac{\partial g}{\partial t} \bigg) \\
\nonumber
 &+& i \frac{\Sigma}{r}(\cos\phi - \sin\phi)\bigg( \frac{\partial f}{\partial t} \frac{\partial g}{\partial \phi} - \frac{\partial f}{\partial \phi} \frac{\partial g}{\partial t} \bigg) \\
 \label{4.2}
&+& i \frac{\Theta}{r}\bigg( \frac{\partial f}{\partial r} \frac{\partial g}{\partial \phi} - \frac{\partial f}{\partial \phi} \frac{\partial g}{\partial r} \bigg).
\end{eqnarray}
In accordance with (\ref{4.2}), the noncommutativity in the polar coordinate is shown as
\begin{eqnarray}
\label{4.3}
[r,\phi]_\star = i \frac{\Theta}{r},
\end{eqnarray}
similarly to the result of the literature\cite{6,7,8}.

Equation (\ref{4.3}) does not immediately reveal that the generalized Moyal product links Cartesian coordinate system with polar one.
Thus, we examine the consistency by focusing on the fact that the noncommutativity in the Cartesian coordinate $[x,y]_\star = i\Theta$ is the relation of orthogonal lengths.
To compare the polar noncommutativity with the Cartesian one, we choose the radial length $r$ and the circumferential length $r\phi$ as a combination of orthogonal lengths in the polar coordinate.
Then in a three-dimensional spacetime, the commutation relation between $r$ and $r\phi$ is
\begin{eqnarray}
\label{4.4}
[r, r\phi]_\star = i\Theta,
\end{eqnarray}
which is determined only by $\Theta$ similarly to the commutation relation in the Cartesian coordinate system.
Therefore, we can say that the generalized Moyal product gives equivalent noncommutativity in both coordinate systems.

Naively we could possibly regard the consistency of the commutation relation between two distinct coordinate systems is obtained when the commutator does not depend on any variables such as $\theta$ or $\phi$.
While the above criterion may seem to be trivially satisfied, it is not the case in four or more dimensional spacetimes.
To see this property explicitly, let us investigate noncommutativity in the four-dimensional polar coordinate $(t, r, \theta, \phi)$ defined by vielbein
\begin{eqnarray}
\label{5.1}
\begin{split}
e^a{_{\mu}} &= \left(
\begin{array}{cccc}
1 & 0 & 0 & 0 \\
0 &  \sin\theta \cos \phi  & r \cos\theta \cos\phi & -r\sin\theta \sin \phi \\
0 & \sin \theta \sin \phi & r\cos\theta \sin \phi & r\sin\theta \cos\phi \\
0 & \cos\theta & -r \sin\theta & 0 \\
\end{array} 
\right)
\
,
\\
e_a{^{\mu}} &= \left(
\begin{array}{cccc}
1 & 0 & 0 & 0 \\
0 & \sin\theta \cos \phi  &  \cos\theta \cos\phi /r & - \sin\phi/r\sin\theta \\
0 & \sin \theta \sin \phi & \cos\theta \sin \phi /r &  \cos\phi /r\sin\theta \\
0 & \cos\theta & - \sin\theta /r & 0 \\
\end{array} 
\right).
\
\end{split}
\end{eqnarray}
In this spacetime, the commutation relation between arbitrary functions $f(t, r,\theta,\phi)$ and $g(t, r,\theta,\phi)$ is derived as
\begin{strip}
\hrulefill{\hspace{90mm}}
\begin{eqnarray}
\nonumber
[f, g]_\star &=& 
i \Sigma(\sin\theta\cos\phi + \sin\theta\sin\phi + \cos\theta) \bigg( \frac{\partial f}{\partial t}\frac{\partial g}{\partial r} - \frac{\partial f}{\partial r}\frac{\partial g}{\partial t}\bigg) + i \frac{ \Sigma}{r}(\cos\theta\cos\phi + \cos\theta\sin\phi - \sin\theta) \bigg( \frac{\partial f}{\partial t}\frac{\partial g}{\partial \theta} - \frac{\partial f}{\partial \theta}\frac{\partial g}{\partial t}\bigg) \\
\nonumber
&+& i \frac{\Sigma}{r\sin\theta}(\cos\phi - \sin\phi) \bigg( \frac{\partial f}{\partial t}\frac{\partial g}{\partial \phi} - \frac{\partial f}{\partial \phi}\frac{\partial g}{\partial t}\bigg) + i \frac{\Theta}{r}\bigg[ 1 - \frac{\cos\theta}{\sin\theta}(\cos\phi+\sin\phi) \bigg] \bigg( \frac{\partial f}{\partial r}\frac{\partial g}{\partial \phi} - \frac{\partial f}{\partial \phi}\frac{\partial g}{\partial r}\bigg) \\
\label{5.2}
&+& i \frac{\Theta}{r}(\cos\phi - \sin\phi) \bigg( \frac{\partial f}{\partial r}\frac{\partial g}{\partial \theta} - \frac{\partial f}{\partial \theta}\frac{\partial g}{\partial r}\bigg) + i \frac{\Theta}{r^2}\bigg( \frac{\cos\theta}{\sin\theta}+ \cos\phi + \sin\phi\bigg)\bigg( \frac{\partial f}{\partial \theta}\frac{\partial g}{\partial \phi} - \frac{\partial f}{\partial \phi}\frac{\partial g}{\partial \theta}\bigg),
\end{eqnarray}
\hspace{90mm}\hrulefill
\end{strip}
\hspace{-1mm}by which we can obtain naturally usual commutation relations (\ref{107}) when substituting $x=r\sin\theta\cos\phi$, $y=r\sin\theta\sin\phi$ and $z=r\cos\theta$.

This time, we employ the radial length $r$, the meridian length $r\theta$ and the latitudinal length $r\phi \sin\theta$ as orthogonal quantities in the polar coordinate in order to compare them with $x$, $y$ and $z$.
In contrast to the case of the Cartesian coordinate, in which the noncommutativity is determined only by a constant as in (\ref{107}), the commutation relation contains inevitably $\theta$ and $\phi$ in the polar coordinate.
Actually, commutation relations are
\vspace{-1mm}
\begin{eqnarray}
\label{5.3}
[r, r \theta]_\star &=&  i \Theta(\cos\phi - \sin\phi), \\
\nonumber
[r, r \phi \sin\theta]_\star &=& i\Theta \big[ \sin\theta - \cos \theta (\cos\phi + \sin\phi) \\
\label{5.4}
&& \hspace{10mm} +\phi\cos\theta( \cos\phi - \sin\phi) \big], 
\end{eqnarray}
\begin{eqnarray}
\nonumber
[r\theta, r\phi \sin\theta]_\star &=& i\Theta\big[ \theta\sin\theta - \theta\cos\theta(\cos\phi + \sin\phi) \\
\nonumber
&& \hspace{8mm} +\theta\phi \cos\theta(\cos\phi - \sin\phi) \\
\nonumber
&& \hspace{8mm}- \phi \sin\theta(\cos\phi-\sin\phi) \\
\label{5.5}
&& \hspace{8mm}+ \cos\theta + \sin\theta(\cos\phi + \sin\phi) \big],
\end{eqnarray}
in which, depending on $\theta$ and $\phi$, right hand sides can be zero or larger than $\Theta$.
However such apparently strange situation is not caused by the selection of polar coordinate.
For example, in the Cartesian coordinate where (\ref{107}) is defined, the quantities $X=(x-y)/\sqrt{2}$ and $Y=(x+y)/\sqrt{2}$ satisfy $[X, z]_\star = -i\sqrt{2}\Theta$ and $[Y, z]_\star = 0$.
That is, even in the Cartesian coordinate, inhomogeneous and anisotropic noncommutativity can be realized by simple coordinate axis rotation.
Obviously this means that our starting coordinate system inevitably determines ``standard frame'' with which the commutators should be homogeneous.
Although it would be quite interesting if the noncommutative spacetime itself has a particular direction related with the quantum property by nature, such an alternative interpretation does not hold since the difference between $(x, y)$ and $(X, Y)$ is not essential, but just artificial.

The apparent inconsistency that the commutation relation in the polar coordinate is complicatedly dependent on coordinate components is merely due to the fact that $(r, \theta, \phi)$ are different from $(x, y, z)$ in  the properties of each elements.
Actually, unlike $x$, $y$ or $z$, which has a fixed direction in the three-dimensional space, $r$-, $\theta$- or $\phi$-axises can have various directions.
Hence, in order to examine the commutation relation of, say,  $[r, r\theta]_\star$ is consistent with the corresponding one in the Cartesian coordinate, it would be sufficient to show their consistency for the particular choice of $r$ and $r\theta$ because it could be generalized to an arbitrary set of $r$ and $r\theta$ with any other $\theta$ and $\phi$ by appropriately redefining the corresponding combination of $x$, $y$ and $z$.
Here, for simplicity, the consistencies with three cases $[x, y]_\star$, $[y, z]_\star$ and $[z, x]_\star$ in the Cartesian coordinate will be specified.
Then, we can show that the commutation relations (\ref{5.3})--(\ref{5.5}) are determined solely by a constant.

First, we compare the commutation relation of the Cartesian coordinate with (\ref{5.3}) which can be regarded as a noncommutativity $[z, x]_\star$ when $\phi=0$, and is shown as 
\begin{eqnarray}
\label{5.6}
[r, r \theta]_\star \Big|_{\phi=0} = i\Theta = [z, x]_\star.
\end{eqnarray}
In the case of $\phi=-\pi/2$, identically to noncommutativity of $y$ and $z$, (\ref{5.3}) becomes
\begin{eqnarray}
\label{5.7}
[r, r \theta]_\star \Big|_{\phi=-\pi/2} = i\Theta = [y, z]_\star.
\end{eqnarray}
According to (\ref{5.6}) and (\ref{5.7}), the commutation relations in the polar coordinate are determined only by  $\Theta$ for the suitable pairs of direction.
In other words, the commutation relations on the four-dimensional spacetime in polar coordinate as (\ref{5.3})--(\ref{5.5}) are just seemingly complicated because they are composed of multiple noncommutativity in Cartesian coordinate.

Next, when examining the commutation relation between $r$ and $ r \phi \sin \theta$, the relation can be simplified by setting $\theta$ to be constant as $\theta = \theta_{const}$.
Thus, the commutation relation derived from (\ref{5.2}) is
\begin{eqnarray}
\label{5.8}
\begin{split}
[r, r\phi& \sin\theta_{const}]_\star \\
&= i\Theta  \big[ \sin\theta_{const} - \cos \theta_{const} (\cos\phi + \sin\phi)\big].
\end{split}
\end{eqnarray}
Furthermore, by setting $\theta_{const}=\pi/2$ for comparison, (\ref{5.8}) is expressed as
\begin{eqnarray}
\label{5.9}
[r, r\phi \sin\theta_{const}]_\star  \Big|_{\theta_{const}=\pi/2} = i\Theta = [x, y]_\star,
\end{eqnarray}
which is also consistent with the commutation relation of  $x$ and $y$ in the three-dimensional case (\ref{4.4}).
The point we have to be careful when evaluating noncommutativity associated with $z$-direction is that $\phi$ is not defined when $\theta_{const}=0$.
Thus we consider the situation $\theta_{const}\ll1$, that is,
\begin{eqnarray}
\label{5.9}
[r, r\phi \sin\theta_{const}]_\star  \Big|_{\theta_{const}\ll1} \cong - i\Theta (\cos\phi + \sin\phi),
\end{eqnarray}
which is expressed in the same form as (\ref{5.3}). 
Therefore $[r, r\phi \sin\theta]_\star$ is also equivalent to the commutation relation in the Cartesian coordinate system, such as
\begin{eqnarray}
\begin{split}
  [r, r\phi \sin \theta_{const}&]_\star  \Big|_{\theta_{const}\ll1} \\
                       &= i\Theta  = \begin{cases} 
                                          [z, x]_\star  &{\rm for}\ \phi = \pi  \\
                                          [y, z]_\star  &{\rm for}\ \phi = -\pi/2 
                                           \end{cases}.
\end{split}                                           
\end{eqnarray}

Finally, we investigate the commutation relation between the meridian length $ r \theta $ and the parallel length $ r \phi \sin \theta $.
Here, in order to clarify this relation, the radial length is set to be constant as $ r = r_{const} $.
Then, the commutation relation is derived as 
\begin{eqnarray}
\label{5.12}
\begin{split}
[r_{const}\theta&, r_{const}\phi \sin\theta]_\star \\
       &= i\Theta \Big[\cos\theta + \sin\theta(\cos\phi + \sin\phi) \Big],
\end{split}
\end{eqnarray}
in accordance with (\ref{5.2}).
The explicit form of noncommutativity related to $x$- and $y$-directions can be written down by applying $\theta \ll 1$ as
\begin{eqnarray}
\label{5.10}
[r_{const}\theta, r_{const}\phi \sin\theta]_\star \Big|_{\theta \ll 1} \cong i\Theta = [x,y]_\star,
\end{eqnarray}
using (\ref{5.12}).
Further, in the case of $\theta=\pi/2$, the commutation relation is derived as
\begin{eqnarray}
\label{5.11}
[r_{const}\theta, r_{const}\phi \sin\theta]_\star \Big|_{\theta=\pi/2} = i\Theta (\cos\phi + \sin\phi),
\end{eqnarray}
which is also qualitatively equivalent to (\ref{5.3}) and (\ref{5.9}), and thus it coincides with $[y,z]_\star$ and $[z,x]_\star$ in each $\phi$-angle, such as
\begin{eqnarray}
\begin{split}
  [r_{const}\theta, r_{const}&\phi \sin\theta]_\star \Big|_{\theta=\pi/2} \\
                                &= i\Theta = \begin{cases} 
                                                   [y, z]_\star  &{\rm for} \phi = 0  \\
                                                   [z,x]_\star  &{\rm for} \phi = \pi/2
                                                  \end{cases} .
\end{split}
\end{eqnarray}

Before the consistency of four-dimensional spacetime can be shown and investigated here, noncommutativity could not even been defined in the three-dimensional polar coordinate.
The above results indicate that, at the appropriate $\theta$- and $\phi$-angles, the commutation relation in the four-dimensional spacetime in polar coordinate is also determined substantially by the constant, $\Theta$.
Hence, we  can say that the generalized Moyal product with $\theta^{\mu\nu}$ defined as equation (\ref{103}) provides consistent noncommutativity in Cartesian and polar coordinate systems.
This is an important result in the field of gravity dealing with various coordinate systems, which we will see in the next section.

\section{\label{sec4}Noncommutativity in axisymmetric spacetime}
The discussion of the previous section revealed that the generalized Moyal product is also useful in polar coordinate systems not only in Cartesian coordinate.
That kind of operation, the generalized Moyal product can be applied to spacetime with gravity by selecting the appropriate vielbein.
In this section, we investigate how the stationary limit surface of axisymmetric spacetime could be modified to examine the effect of spacetime noncommutativity.

In classical spacetime, the axisymmetric solution is the classical Kerr metric

\begin{eqnarray}
\nonumber
g_{00} &=& - \bigg( 1- \frac{2Mr}{\rho^2} \bigg) , \hspace{5mm} g_{11} = \frac{\rho^2}{\Delta}, \hspace{5mm} g_{22} = \rho^2, \\
\nonumber
g_{33} &=& \bigg( r^2 + a^2 + \frac{2Mra^2}{\rho^2}\sin^2\theta \bigg)\sin^2\theta, \\
\nonumber
g_{03} &=& g_{30} = \frac{2Mra}{\rho^2}\sin^2\theta, \\
\nonumber
\rho^2 &=& r^2 + a^2\cos^2\theta, \hspace{5mm} \Delta = r^2 - 2Mr + a^2, \\
\nonumber
\zeta &=& g_{03} / \sqrt{-g_{00}}, \hspace{5mm} \beta^2 = \zeta^2 + g_{33},
\end{eqnarray}
where $M$ and  $J$ are the mass and the the angular momentum of the source respectively and $a = J/M$ is the angular momentum per unit mass.
Also, the corresponding vielbein for the classical Kerr metric are

\begin{strip}
\hrulefill\hspace{90mm}
\begin{eqnarray}
\label{201}
e^a{_{\mu}} &= \left(
\begin{array}{cccc}
\sqrt{-g_{00}} & 0  & 0 & \zeta \\
0 & \sqrt{g_{11}}\sin \theta \cos \phi  & \sqrt{g_{22}}\cos \theta \cos \phi & - \beta \sin \phi \\
0 & \sqrt{g_{11}}\sin \theta \sin \phi & \sqrt{g_{22}}\cos \theta \sin \phi & \beta \cos \phi \\
0 & \sqrt{g_{11}}\cos \theta & - \sqrt{g_{22}}\sin \theta & 0 \\
\end{array} 
\right)
\
,
\\
\label{205}
e_a{^{\mu}} &= \left(
\begin{array}{cccc}
1/\sqrt{-g_{00}} & 0  & 0 & 0 \\
-\beta g^{03} \sin\phi & \sin \theta \cos \phi / \sqrt{g_{11}}  & \cos \theta \cos \phi / \sqrt{g_{22}} & - \sin \phi / \beta \\
\beta g^{03} \cos\phi & \sin \theta \sin \phi / \sqrt{g_{11}} & \cos \theta \sin \phi / \sqrt{g_{22}} &  \cos \phi /\beta \\
0 & \cos \theta / \sqrt{g_{11}} & - \sin \theta / \sqrt{g_{22}} & 0 \\
\end{array} 
\right).
\
\end{eqnarray}
\hspace{90mm}\hrulefill
\end{strip}
In the classical Kerr spacetime, the radius of the stationary limit surface $r_{c.lim}(\theta)$ satisfies $g_{00}(r_{c.lim})=0$.
Especially on the plane of $\theta = \pi/2$, $r_{c.lim}(\pi/2) = 2M$, while in the case of $\theta=0$, $r_{c.lim}(0) = M + \sqrt{M^2 - a^2}$.
The shape of the stationary limit surface depends on the value of $M^2-a^2$, and the oblateness of the surface increases as $a = J/M$ raises.
If we focus on the oblateness of the stationary limit surface, larger $J$ has an identical effect to smaller $M$.
Therefore, it can be naively interpreted that the increase in the oblateness of the stationary limit surface is due to the reduction of gravity.

In this work, the real metric tensor in noncommutative spacetime is defined in a simple form as

\begin{eqnarray}
\label{202}
{\hat g}_{\mu\nu} 
= \frac{1}{2}\eta_{ab}(e^a{_{\mu}} \star e^b{_{\nu}} + e^b{_{\nu}} \star e^a{_{\mu}}).
\end{eqnarray}
Note that since ${\hat g}_{\mu\nu}$ is defined using a generalized Moyal product and a classical vielbein $e^a{_{\mu}}$, the zero-order term in the series expansion of ${\hat g}_{\mu\nu}$ in the noncommutative parameter, is the classical metric $g_{\mu\nu}$.
Here, we assume the radius of the noncommutative stationary limit surface $r_{n.lim}(\theta)$ in the noncommutative Kerr spacetime satisfies ${\hat g}_{00}(r_{n.lim})=0$.
In order to evaluate $r_{n.lim}(\theta)$, it is not necessary to know all the components of ${\hat g}_{\mu\nu}$, and it is sufficient to derive only ${\hat g}_{00}$.
Substituting (\ref{201})--(\ref{205}) into (\ref{202}), ${\hat g}_{00}$ is expressed as

\begin{strip}
\hrulefill\hspace{90mm}
\begin{eqnarray*}
\nonumber
\begin{split}
{\hat g}_{00} = g_{00} 
 +&  \frac{1}{2}\Theta^2 (-g_{00})^{-5/2}(g_{11})^{-1}(g_{22})^{-1} \frac{M^3ra^2(\cos\phi-\sin\phi)^2}{(r^2+a^2\cos^2\theta)^{6}} \\ 
&  \times \bigg[ -(\cos^2 \theta - \sin^2 \theta)(r^2-a^2\cos^2 \theta)^2 + 2a^2 \sin^2 \theta \cos^2 \theta \Big\{ r^2(3-4\alpha) -a^2\cos^2\theta  \Big\}\bigg] \\ 
 +&  \Theta^2 (-g_{00})^{-2}(g_{11})^{-1}(g_{22})^{-1}\frac{M^2a^2(\cos\phi-\sin\phi)^2}{(r^2+a^2\cos^2\theta)^{4}} \\ 
&  \times \Bigg[ r^2(\cos^2\theta- \sin^2\theta)( 1 - 2 \alpha  ) +a^2 \sin^2\theta \cos^2\theta \bigg\{ 1 + 4\alpha(1+\alpha) + \frac{4r^2 (1-2\alpha)}{r^2+a^2\cos^2\theta} \bigg\} \Bigg] \\
\end{split}
\end{eqnarray*}
\end{strip}
\begin{strip}
\begin{eqnarray*}
\begin{split}
 \hspace{16mm}+&  \frac{1}{3}\Theta^2 (-g_{00})^{-3/2}(g_{11})^{-1}(g_{22})^{-1}\frac{M^2a^2(\cos\phi-\sin\phi)^2}{(r^2+a^2\cos^2\theta)^{4}} \\
&  \times \Bigg[ (\cos^2\theta-\sin^2\theta) \Bigg\{ 2\alpha r^2 + (r^2-a^2\cos^2\theta) \frac{1-M/r}{ 1 - 2M/r + a^2/r^2} \Bigg\}  \\
& \hspace{5mm} - 2a^2\sin^2\theta\cos^2\theta \Bigg\{ 4r^2(1-\alpha) (r^2+a^2\cos^2\theta)^{-1}+ \alpha(1+2\alpha) -  \frac{1-M/r}{ 1 - 2M/r + a^2/r^2} \Bigg\} \Bigg] \\
 +&  \frac{1}{3}\Theta^2 (-g_{00})^{-3/2}(g_{11})^{-1}(g_{22})^{-1/2}\beta^{-1}\frac{M^2a^2}{(r^2+a^2\cos^2\theta)^{4}} \\
&  \times \sin\theta\cos\theta \Big\{ \sin\theta(\cos\phi+\sin\phi)-2\cos\theta\sin\phi\cos\phi-\cos\theta \Big\}\Big\{ r^2(3-4\alpha) -a^2\cos^2\theta +2\alpha(r^2-a^2\cos^2\theta)\Big\} \\
 +&  \frac{1}{3}\Theta^2 (-g_{00})^{-3/2}(g_{11})^{-1/2}(g_{22})^{-1}\beta^{-1} \frac{M^2a^2}{(r^2+a^2\cos^2\theta)^{4}} \\
 \label{110}
&  \times \Big\{ \cos\theta(\cos\phi+\sin\phi)+2\sin\theta\sin\phi\cos\phi+\sin\theta \Big\} \Big\{ r(\cos^2\theta-\sin^2\theta)(r^2-a^2\cos^2\theta) -2a^2r\sin^2\theta\cos^2\theta \Big\},
\end{split}
\end{eqnarray*}
\hspace{90mm}\hrulefill
\end{strip}
where $ \alpha = (r^2 -a^2\cos^2 \theta)(r^2 +a^2\cos^2 \theta)^{-1}$.
The above expression is too lengthy to understand how spacetime noncommutativity works at a glance.
However, in the vicinity of classical stationary limit surface $g_{00}\sim0$, the most dominant contribution must come from the term $(-g_{00})^{-5/2}$.

For simplicity we consider ${\hat g}_{00}$ on the plane of $\theta = \pi/2$ then it can be calculated as
\begin{eqnarray}
\label{111}
{\hat g}_{00} \bigg(\theta=\frac{\pi}{2}\bigg)  \cong - \bigg( 1- \frac{2M}{r}\bigg) + \Xi,
\end{eqnarray}
\begin{eqnarray}
\nonumber
\Xi &\equiv&  \frac{1}{2}\Theta^2 \bigg( 1- \frac{2M}{r}\bigg)^{-5/2} \\
\label{112}
      &&\hspace{5mm}\times \bigg( 1-\frac{2M}{r}+\frac{a^2}{r^2}\bigg)(\cos\phi - \sin\phi)^2\frac{M^3a^2}{r^9},
\end{eqnarray}
where $\Xi$, which is responsible for the effect of noncommutativity on the equatorial plane, is always non-negative i.e. $\Xi \geq 0$ and has an opposite sign to the classical term, $-(1-2M/r)$.
The stationary limit radius $r_{n.lim}(\theta)$ that satisfies ${\hat g}_{00}(r_{n.lim})=0$ on the equatorial plane is 
\begin{eqnarray}
\label{115}
r_{n.lim}\bigg(\theta=\frac{\pi}{2}\bigg) \cong \frac{2M}{1-\Xi},
\end{eqnarray}
so it is clear that  the noncommutative stationary limit surface has a larger size than a classical stationary limit surface, that is $r_{n.lim} > r_{c.lim}(\theta =\pi/2)$.

On the other hand, since $\phi$ cannot be defined when $\theta=0$, the effect of noncommutativity in the axial direction is examined with $\theta \ll 1$.
At this time, ${\hat g}_{00}$ in the axial direction is derived as
\begin{eqnarray}
\label{113}
{\hat g}_{00} (\theta \ll 1)  \cong - \bigg( 1- \frac{2Mr}{r^2 + a^2}\bigg) + \Gamma,
\end{eqnarray}
\begin{eqnarray}
\nonumber
\Gamma &\equiv&  -\frac{1}{2}\Theta^2 \bigg( 1- \frac{2Mr}{r^2 + a^2}\bigg)^{-3/2}\\
\label{114}
&&\hspace{5mm}\times (\cos\phi - \sin\phi)^2 \frac{(r^2-a^2)^2}{(r^2+a^2)^7}M^3ra^2,
\end{eqnarray}
where $\Gamma$, which corresponds to the effect of noncommutativity in the axial direction, is always non-positive i.e. $\Gamma \leq 0$ and has the same sign as the classical term.
Therefore,  the radius of stationary limit surface in the axial direction obtained from (\ref{113}) is expressed as 
\begin{eqnarray}
\label{116}
r_{n.lim}(\theta \ll 1) \cong \frac{M}{1-\Gamma} + \sqrt{\bigg(\frac{M}{1-\Gamma} \bigg)^2 - a^2},
\end{eqnarray}
so the relation $r_{n.lim}<r_{c.lim}(\theta \ll 1)$ is established in the axial direction.

As shown, the effect of noncommutativity causes the stationary limit surface to expand in the equator direction and contract in the axial direction, so that the oblateness of the surface is increased.
Classically, the increase in the oblateness of the surface could be regarded as the reduction of gravity.
From this result, we can naively interpret that in the noncommutative spacetime, conventional gravity is mitigated.
Similarly, previous studies \cite{16, 23, 24} have suggested that the effect of noncommutativity acts in the opposite direction to attractive gravity.
Therefore, regarding the effect of noncommutative spacetime, we can consider that our research has obtained consistent results with previous research.

\vspace{5mm}
\section{\label{sec7}Conclusions}
In this paper, we showed that the generalized Moyal product is useful not only in the Cartesian coordinate but also in any coordinate system.
In particular, we focused on noncommutativity in the polar coordinate and derived commutation relations in three and four-dimensional spacetimes.
Thus, we confirmed that the commutation relations in Cartesian and polar coordinates are consistent by using the generalized Moyal product.
Also, in the three-dimensional spacetime in polar coordinate, the commutation relation derived in this way was equivalent to that in the previous works \cite{6,7,8}.
Moreover, since the generalized Moyal product is applicable in arbitrary spacetime, as an example we investigated the effect of noncommutativity in the noncommutative Kerr spacetime.
As a result it was demonstrated that the time component of the metric near the stationary limit surface of the noncommutative Kerr spacetime is expressed as (\ref{111}) in the equatorial plane and (\ref{113}) in the axial direction.
Using these formulae, we have shown that the effect of noncommutativity enhances the oblateness of the stationary limit surface in the axisymmetric spacetime.
From the results, it is indicated that the effect of gravity is alleviated in the noncommutative spacetime.
This is consistent with the result obtained in the previous research on the noncommutative gravitational theory \cite{16, 23, 24}, and it is a significant indication that the similar interpretation can be derived by a different approach in this study.

A notable point is that there is a clear difference in the property of noncommutativity between three-dimensional and four-dimensional spacetime.
In a three-dimensional spacetime, the commutation relation is defined only by $\Theta$ without any coordinate angles as you can see in $[x,y]_\star=i\Theta$ and (\ref{4.4}).
It means that the noncommutativity is homogeneous and isotropic throughout the space.
However, in the four-dimensional spacetime, the commutation relation can be inhomogeneous and anisotropic depending on the angles $\theta$  and $\phi$ as  (\ref {5.3})--(\ref{5.5}), and it could be even commutable.
This is caused by the coincident cancellation of independent noncommutativities $[x,y]_\star$, $[y,z]_\star$ and $[z,x]_\star$ with each other.
Therefore, it can be considered that nonconstant commutation relation dependent on the position is a specific behaviour of the four or more dimensional noncommutative spacetimes.
That is, in the four or more dimensional noncommutative spacetime, the effect of noncommutativity might be observed differently depending on the particular choice of the coordinate system of the observer.
It is similar to the well-known fact in quantum mechanics that the direction in which the uncertainty of, say, the angular momentum remains is determined by the observation.
If we can describe how such a distinct feature can be detected in the quantum gravitational events, we can verify our formularization by future experiments.
Although the theoretical issues to be solved in constructing the noncommutative gravitational theory still remain, some of various models proposed in the future based on the generalized Moyal product we have presented in this paper would contribute to the development of quantum gravity.

\begin{acknowledgements}
Ryouta Matsuyama thanks Yoneda Yoshimori Education Scholarship in Kanagawa University for the grant that made it possible to complete this study.
\end{acknowledgements}




%
%

\end{document}